\documentclass[twocolumn,twoside,slac_two]{revtex4}
\usepackage{graphicx}
\usepackage{fancyhdr}
\newcommand{\numutonue}{$\nu_{\mu}\rightarrow\nu_{e}$ }
\newcommand{\antinumutonue}{$\overline{\nu}_{\mu}\rightarrow\overline{\nu}_{e}$ }

\newcommand{\numu}{$\nu_{\mu}$}
\newcommand{\numubar}{$\overline{\nu}_{\mu}$}
\newcommand{\nue}{$\nu_{e}$}
\newcommand{\nuebar}{$\overline{\nu}_{e}$}
\newcommand{\dm}{{\Delta}m^2}
\newcommand{\ssq}{\sin^{2}2{\theta}}

\newcommand{\dmsq}{$|\Delta m^{2}_{32}|$}

\newcommand{\sintwo}{$\rm sin^{2}(2\theta_{23})$}

\newcommand{\ket}[1]{|#1\rangle}

\begin{document}

\title{Neutrino Oscillation Results from MINOS and MiniBooNE}

\author{Tobias~M.~Raufer}
\affiliation{STFC Rutherford Appleton Laboratory, UK}

\begin{abstract}
  After a brief introduction to neutrino oscillations and a review of
  the world knowledge of neutrino oscillation parameters, we introduce
  two current neutrino oscillation experiments, MINOS\footnote{MINOS:
    Main Injector Neutrino Oscillation Search} and
  MiniBooNE\footnote{BooNE: Booster Neutrino Experiment}. MINOS makes
  precise measurements of the oscillation parameters \dmsq and
  \sintwo. MiniBooNE tests neutrino oscillations in the parameter
  region reported by the LSND experiment, which would require a new
  neutrino state. We review recent experimental results from both
  experiments and give an outlook on future measurements.
\end{abstract}

\maketitle

\thispagestyle{fancy}


\section{Introduction}
Neutrino oscillations are now a well established phenomenon. They
follow from the non-zero and non-degenerate masses of the neutrino
states and from the fact that mass and interaction eigenstates of
the neutrino are not identical. The two sets of eigenstates are
related by a unitary transformation, the PMNS matrix:
\begin{equation}
\ket{\nu_{\alpha}} = \sum_i U^{*}_{\alpha i} \ket{\nu_{i}}
\end{equation}

Several experiments have observed neutrino oscillations and
constrained the entries in the PMNS matrix, commonly parametrised by
three rotation angles $\theta_{ij}$ and a complex phase $\delta$. The
Super-Kamiokande (SuperK) experiment in Japan measured the
disappearance of neutrinos produced in the upper atmosphere dependent
on neutrino flavor, energy and zenith angle \cite{SuperK}, pointing
to the characteristic $L/E$ signature of neutrino oscillations. The
K2K experiment confirmed this result \cite{K2K} using an accelerator
neutrino beam measured at a distance of 250\,km. Neutrino oscillations
have also been measured in solar neutrinos. After a long list of
experiments starting as early as the 1960s, the most precise
measurements of solar neutrino oscillations comes from the Sudbury
Neutrino Observatory (SNO) \cite{SNO}. Finally, measurements at
nuclear reactors have established the smallness of the third mixing
angle, $\theta_{13}$. The best limit on this parameter comes from the
CHOOZ \cite{chooz} experiment in France.

The world knowledge on neutrino oscillation parameters from a global
fit to all available experimental data is shown in Table~\ref{global
  fit} (taken from \cite{MaltoniSchwetzValle}). The final column shows
the experiment with the biggest contribution to a given parameter. It
is clear from this table that neutrino physics has entered the realm
of precision measurements.

\begin{table}[ht]
\begin{center}
\caption{Global fit neutrino oscillation parameters}
\begin{tabular}{|l|c|c|}
  \hline \textbf{Parameter} & \textbf{Best fit $\pm 1\sigma$} & 
  \textbf{Experiment}  \\
  \hline  
  $\Delta m^2_{21}$ & $(7.6 \pm 0.2)\times 10^{-5}\rm{eV^2}$  & KamLAND \\
  \hline $\sin^2\theta_{12}$ & $0.320 \pm 0.023$ &  SNO \\
  \hline \dmsq & $(2.4\pm0.15)\times10^{-3}\rm{eV^2}$ &  MINOS \\
  \hline \sintwo & $0.50 \pm 0.063$ & SK atm \\
  \hline $\sin^2\theta_{13}$ & $< 0.05$ at $3\sigma$ & CHOOZ \\
  \hline
\end{tabular}
\label{global fit}
\end{center}
\end{table}
\subsection{Sterile Neutrinos}
The number of active neutrinos, measured by the LEP experiments using
the width of the $Z^0$ resonance \cite{LEP}, is consistent with three
at high precision. There is however no limit on the number of
non-interacting neutrinos, customarily called sterile neutrinos.

One motivation for the search for sterile neutrinos is a result by the
LSND experiment \cite{LSND} which claimed evidence for \antinumutonue
oscillations with a mass splitting incompatible with the experimental
results given in the previous section.

The MiniBooNE experiment was designed specifically to test this
so-called ``LSND anomaly''.

MINOS also investigates sterile neutrino models by measuring neutral
current neutrino interactions.

\section{The MINOS experiment}
MINOS is a long-baseline neutrino oscillation experiment at Fermilab's
``Neutrinos at the Main Injector'' (NuMI) beam line. It utilises two
detectors situated at distances of $\sim$1\,km and $\sim$735\,km from
the neutrino production target respectively. The near detector
measures the beam composition and energy spectrum, making use of the
large number of neutrino interactions close to the source. This is
compared to the measurement at the far detector in order to study
neutrino oscillations.

\subsection{The NuMI beam}
The NuMI beam line uses 120\,GeV protons from the Main Injector
accelerator incident on a segmented graphite target where they produce
mainly $\pi$- and $K$-mesons. These secondary particles are focused by
two parabolic horns and subsequently decay in a 675\,m long
helium-filled decay volume to produce neutrinos. The remaining hadrons
are absorbed in a water-cooled beam dump at the end of the decay
volume while muons produced in the decay are absorbed in the rock
separating the near detector hall from the hadron absorber. A small
fraction of muons further decay before being absorbed, producing
electron and anti-muon neutrinos. The resulting neutrino beam has the
following composition: 92.9\% \numu, 5.8\% \numubar, 1.2\% \nue\ and
0.1\% \nuebar\ for the low energy (LE) beam configuration.

The focusing peak and thus the neutrino energy spectrum of the NuMI
beam can be varied by moving the target with respect to the magnetic
horns. This feature is unique to the NuMI beam and is used in all
MINOS analyses in order to reduce systematic uncertainties related to
hadron production in the neutrino production target. 

\subsection{The MINOS detectors}
The MINOS detectors are functionally identical steel/scintillator
sampling calorimeters equipped with a magnetic field to allow charge
sign determination of muon tracks. The far detector has a mass of
5.4\,kt and is situated at a depth of 2130\,m.w.e.\ in the Soudan
Underground Laboratory. The smaller near detector is $\sim$1\,kt and
is situated 100\,m below ground at Fermilab. The detectors are
constructed from 2.5\,cm thick steel planes mounted with a 1\,cm thick
layer of plastic scintillator segmented into $\sim$4\,cm wide
strips. The scintillation light is collected by wavelength shifting
fibres and readout by multi-anode photomultiplier tubes.  

The detectors are calibrated using single particle response data
collected with a smaller 12\,t calibration detector at the CERN PS
test-beam and a light injection system tracking PMT gain
variations. Inter-detector calibration of the energy response is
achieved using cosmic ray muons.

\section{MINOS data analysis}
In this document, we present results from two different analyses and
give a status report on a third one. They can be classified by the
main type of interaction of interest: \numu\ charged current, neutral
current and \nue\ charged current. At the time of writing of this
document, MINOS has collected an integrated $4.8\times10^{20}$
protons-on-target (POT), most of which was recorded in the LE beam
configuration. The results reported here use a data sample of
$2.5\times10^{20}$ POT.

\begin{figure}[ht]
\centering
\includegraphics[width=70mm,clip]{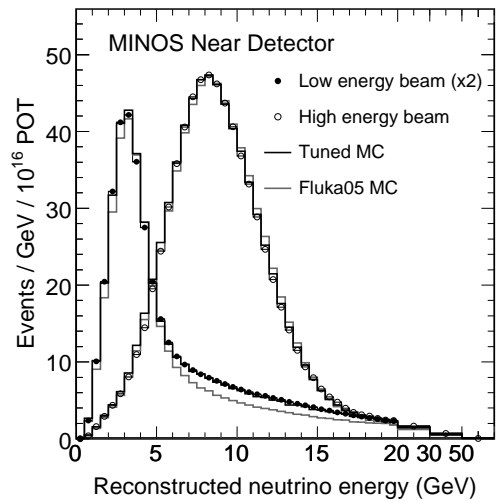}
\caption{Near Detector data and Monte Carlo for the low and high
  energy beam configurations. The grey lines show the untuned Monte
  Carlo simulation based on Fluka05. The black lines show the Monte
  Carlo after hadron production tuning is applied. The agreement is
  much improved.}
  \label{beam_tuning}
\end{figure}

The three analyses have many steps on common. Firstly, they all rely
on a good understanding of the details of the neutrino beam. The major
unknown here is the production of hadrons in the NuMI target. A
detailed Monte Carlo simulation of the beam line components based on
Fluka05 reproduces the data well but not perfectly. To further improve
data/Monte Carlo agreement, the hadron production is parametrised as a
function of $x_F$ and $p_t$ and a simultaneous fit is performed to
the near detector spectra in various different beam
configurations. Individual neutrino interactions are reweighted
according to the $x_F$ and $p_T$ of the parent pion (or kaon). Figure
\ref{beam_tuning} shows near detector data and Monte Carlo for two
different beam configurations before and after the tuning procedure.
A total of seven different beam configurations were used in the fit.
The thus improved hadron production model is then used in the
prediction of the far detector neutrino energy spectrum.

Another common feature is the ``blind analysis'' technique. This
involves not looking at the far detector data until the complete
analysis procedure -- e.g. selection cuts, extrapolation, fitting,
etc.\ -- is defined. The near detector data is not sensitive to
oscillations at the atmospheric scale and was therefore accessible
throughout the development of the analyses, thus allowing to
understand low level detector effects in the data.

\subsection{\numu \ charged current analysis}
\label{MINOSCC}
The charged current disappearance analysis is the main analysis of
MINOS. Measuring the charged current energy spectrum in both
detectors, MINOS measures the neutrino oscillation parameters \dmsq
and \sintwo. The analysis reported here is an update of our initial
results published in \cite{Michael:2006rx}.

Charged current candidate events are selected by demanding a
reconstructed track within $53^\circ$ of the beam direction with a
vertex inside the fiducial volume and in time with the beam spill. To
further reject neutral current background events a likelihood-ratio
discriminant is constructed from six variables, reflecting event
topology, energy loss, inelasticity, etc. A more detailed discussion
of the charged current event selection and extrapolation to the far
detector for this analysis can be found in \cite{CCspring2008}.

\begin{figure}[ht]
\centering
\includegraphics[width=75mm,clip]{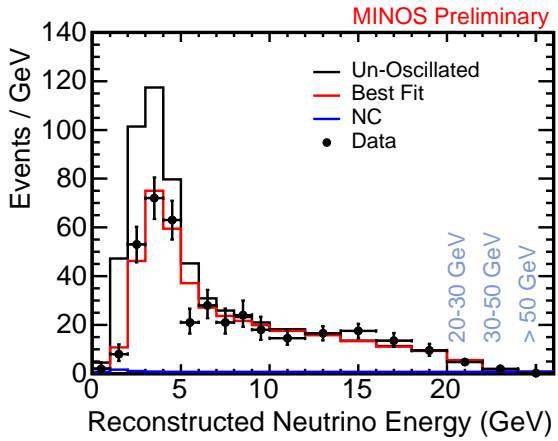}
\caption{Reconstructed far detector neutrino energy spectrum. The
  black markers with error bars show the data, the black and red lines
  show unoscillated and best-fit Monte Carlo predictions. The neutral
  current background is shown in blue.}
  \label{CCspectrum}
\end{figure}

Figure \ref{CCspectrum} shows the selected charged current energy
spectrum corresponding to an exposure of $2.5 \times 10^{20}$ POT. The
points with error bars show the far detector data and the black and
red lines show unoscillated and oscillated Monte Carlo predictions
respectively. 
The best-fit values for the oscillation parameters are: \dmsq $=
2.38^{+0.20}_{-0.16}\times10^{-3}\rm{eV}^{2}$ and
\sintwo$=1.00_{-0.08}$. The errors given combine both statistical and
systematic errors. The latter were included in the fit via nuisance
parameters.

\subsection{Neutral current analysis}
If neutrino oscillations only involve the usual active neutrino
flavours, the neutral current event rate remains unchanged. The
measurement of neutral current interactions therefore allows the
investigation of oscillations into sterile neutrinos, i.e. neutrino
states with no standard model interactions.

Neutral current events in MINOS are selected with a set of simple cuts
using event shape variables. The main distinguishing feature between
charged current and neutral current events is the presence or lack of
a muon in the final state leading to a reconstructed track in the
detector. Short events, events with no track and events where the
reconstructed track is shorter than the hadronic shower are classified
as neutral current.

In the near detector, additional cuts on spacial separation and timing
are applied to make sure the event is well reconstructed and not a
remnant of another bigger event in the vicinity.

\begin{figure}[ht]
\centering
\includegraphics[width=85mm,clip]{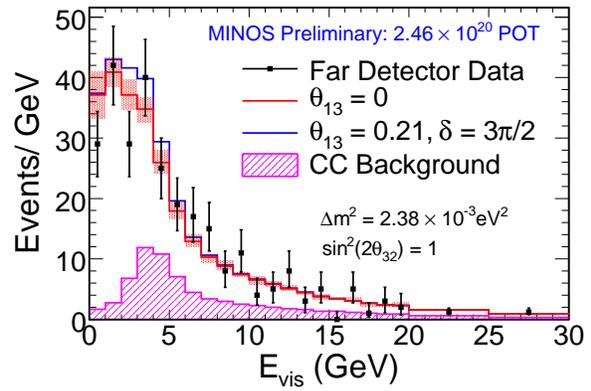}
\caption{Far Detector reconstructed energy spectrum for selected
  neutral current events. The data is shown as black points with error
  bars, the Monte Carlo is shown in blue and red for two extremes of
  the third angle $\theta_{13}$. The charged current background is
  shown in pink.}
  \label{ncSpectrum}
\end{figure}

In the far detector, an additional set of cuts removing cosmic rays
and low energy noise is applied. The resulting neutral current
reconstructed energy spectrum is shown in Figure \ref{ncSpectrum}. 

The Monte Carlo expectation for this spectrum depends on the value of
the third mixing angle $\theta_{13}$ since \nue\ interactions have a
high probability of being classified as neutral currents in this
analysis. We therefore show two Monte Carlo lines, one for
$\theta_{13}=0$ (red) and one for electron neutrino appearance at the
CHOOZ limit (blue). The width of the red band reflects the systematic
uncertainties.  For reconstructed energies below 3\,GeV, a deficit
with respect to the Monte Carlo expectation for $\theta_{13}=0$ of
1.15\,$\sigma$ is observed. The data is thus consistent with no active
neutrino disappearance.

While the above discussion is model independent, it is also
instructive to compare the data to a particular model involving mixing
between active and sterile neutrinos. We use a simple model
introducing a single sterile neutrino state and retaining
one-mass-scale dominance. Assuming no mixing of the first and second
mass eigenstates with the sterile state and assuming that the first
and fourth mass eigenstate are nearly degenerate leads to the
following oscillation probabilities:
\begin{eqnarray}
P_{\nu_\mu\rightarrow\nu_\mu}&=
&1 - 4|U_{\mu3}|^2(1-|U_{\mu3}|^2)\sin^2(1.27\Delta m_{32}^2L/E) \nonumber\\
P_{\nu_\mu\rightarrow\nu_e}&=
&4|U_{\mu3}|^2|U_{e3}|^2\sin^2(1.27\Delta m_{32}^2L/E) \nonumber\\
P_{\nu_\mu\rightarrow\nu_s}&=
&4|U_{\mu3}|^2|U_{s3}|^2\sin^2(1.27\Delta m_{32}^2L/E) \nonumber\\
P_{\nu_\mu\rightarrow\nu_\tau}&=
& 1 - P_{\nu_\mu\rightarrow\nu_\mu} 
- P_{\nu_\mu\rightarrow\nu_e}  - P_{\nu_\mu\rightarrow\nu_s}.  
\end{eqnarray}

The charged and neutral current energy spectra are fitted
simultaneously to this 4-flavour model under two different
scenarios. The first scenario assumes no \nue\ appearance,
i.e. $|U_{e3}|^2=0$, the second assumes \nue\ appearance at the CHOOZ
limit, $|U_{e3}|^2=0.04$. The results are:
\begin{eqnarray}
  |U_{s3}|^{2} = 0.14^{+0.18}_{-0.13}& |U_{\mu3}|^{2} = 0.50^{+0.16}_{-0.15}
  & \mathit{(|U_{e3}|^2 = 0)}\nonumber \\
  |U_{s3}|^{2} = 0.21^{+0.20}_{-0.12}& |U_{\mu3}|^{2} = 0.48^{+0.18}_{-0.12}
  & \mathit{(|U_{e3}|^2 = 0.04)} \nonumber 
\end{eqnarray}
Systematic errors were included in the fit as nuisance parameters; the
errors quoted above thus include both statistical and systematic
uncertainties. The values for $|U_{\mu3}|^2$ are consistent with the
parameter \sintwo\ reported in section \ref{MINOSCC}.

More details on the analysis of neutral current interactions in MINOS
can be found in \cite{Adamson:2008jh}.

\subsection{$\nu_e$ appearance analysis}
The \nue\ appearance analysis investigates the sub-dominant oscillation
channel \numutonue which is sensitive to the third angle in the PMNS
matrix, $\theta_{13}$. The non-observation of the reverse process at
the CHOOZ reactor experiment \cite{chooz} provides an upper limit for
this parameter. A potential observation of $\nu_e$ events in the MINOS
far detector beyond the inherent level in the NuMI beam would provide
evidence for a non-zero value below the CHOOZ limit.

$\nu_e$ events are characterised by an electromagnetic shower in the
detector in addition to the hadronic activity caused by the struck
nucleus. The success of this analysis therefore relies on the ability
to identify electromagnetic showers in a large background of hadronic
activity. Due to the relatively coarse segmentation of the MINOS
detectors, this is a very difficult task.

$\nu_e$ candidates are selected using a neural network
technique. Since these techniques rely on good data/Monte Carlo
agreement, two independent data driven methods using the near detector
have been developed in order to improve the modeling of the
backgrounds. This is possible since the baseline at the near detector
is too short for oscillations to develop.

One method takes well understood, well reconstructed charged current
events and removes the muon track, leaving a sole hadronic
shower. This remnant event is then reconstructed again in order to
determine, how often such an event is misclassified as an
electromagnetic shower. Using this method, a discrepancy of 20\,\% is
found, with the Monte Carlo overestimating the background. A second
method compares data taken with and without the focusing horns being
powered. This allows a deconvolution of the different background
components since the composition is quite different in the two
samples. 

\begin{figure}[ht]
\centering
\includegraphics[width=80mm,clip]{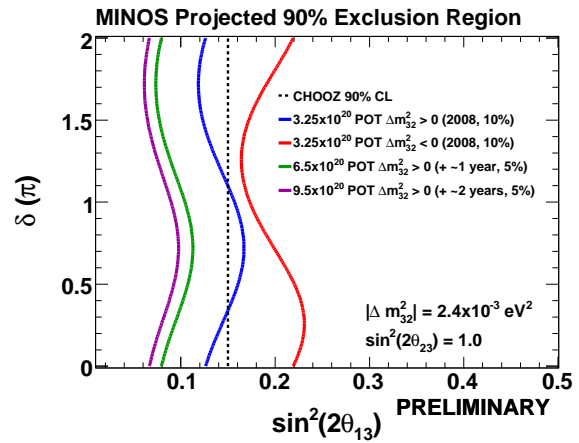}
\caption{Sensitivity to $\theta_{13}$ for three different
  exposures. For our current data set of $3.25\times10^{20}$ the
  sensitivity for both the normal (blue) and inverted (red) mass
  hierarchies are shown. A 10\% systematic error is included in all
  predictions. With increasing statistics, improvements over the
  CHOOZ bound can be made.}
  \label{nueSens}
\end{figure}

The results obtained using the two different methods are consistent,
justifying an ad-hoc correction which is applied to the Monte Carlo
prediction. The background estimates are extrapolated to the far
detector where the sensitivity to measuring $\theta_{13}$ is
evaluated. These sensitivities for three different exposures are shown
in Figure \ref{nueSens}.

The data for this analysis are still blinded and final studies are
being performed, including the investigation of data
``sidebands''. Results from this analysis are eagerly awaited and are
expected later this year.

\section{The MiniBooNE experiment}
MiniBooNE is a short-baseline neutrino oscillation experiment using
Fermilab's booster neutrino beam line. With a source-detector distance of
$\sim$0.5\,km and a peak neutrino energy of 0.8\,GeV it is
specifically designed to test the parameter space of the LSND claim
discussed in section 1.1.

MiniBooNE uses a conventional neutrino beam similar to the NuMI beam
described above. The booster's 8\,GeV protons impinge on a beryllium
target at a rate of up to 4\,Hz. The secondary mesons are focused by a
single pulsed magnetic focusing horn and allowed to decay in a
50\,m-long decay region. The detector is separated from the end of the
decay volume by $\sim$500\,m of earth.  

The MiniBooNE detector consists of a spherical volume filled with 800
tons of mineral oil ($\rm{CH_2}$), acting both as a scintillator and
Cerenkov medium.  The light produced in the detector is read out by
1280 inward facing 8-inch photomultiplier tubes (PMTs) providing a
10\,\% photo-cathode coverage, viewing a target volume 575\,cm in
diameter. The target volume is surrounded by a 35\,cm thick, optically
isolated veto region viewed by 240 PMTs.

\section{MiniBooNE oscillation analysis}
The MiniBooNE analysis was conducted in a blinded fashion. Only Monte
Carlo simulation and data insensitive to \numutonue oscillations was
used to develop the analysis procedure.

As in the MINOS experiment, hadron production uncertainties in the
target are an important contributor to the systematic error for
MiniBooNE analyses. MiniBooNE uses a Sanford-Wang parametrisation to
the pion data from the HARP experiment \cite{harp} and a fit to the
world kaon production data in the range of 10--24\,GeV to improve on
this uncertainty. The quoted uncertainties are 17\,\% for the pion and
30\,\% for the kaon flux. The predicted kaon flux is checked using
off-axis muon counters and high energy events in the MiniBooNE
detector.

To select electron-like neutrino interactions in MiniBooNE, events are
first divided into so-called subevents, i.e. sets of PMT hits which
are clustered within $\sim$100\,ns. Only subevents in time with the
beam spill are selected. Subevents are required to have more than 200
hits in the tank, less than 6 hits in the veto region and a
reconstructed vertex position with $R<500$\,cm, eliminating cosmic ray
muons and associated decay electrons. \nue\ charged current event
candidates a required to have exactly one subevent.

After these initial selection cuts, two visible energy dependent
likelihood ratio discriminants are constructed in order to further
reduce the backgrounds. The first one separates muon-like from
electron-like Cerenkov rings. The second discriminant separates the
remaining events into electron-like events and events from
$\pi^0$-decay. For the complete selection, the Monte Carlo predicts an
efficiency for \nue\ charged current quasi-elastic events of
$20.3\pm0.9\%$. Further details on the event selection can be found in
\cite{miniboone}.

\begin{figure}[t]
\centering
\includegraphics[width=80mm,clip]{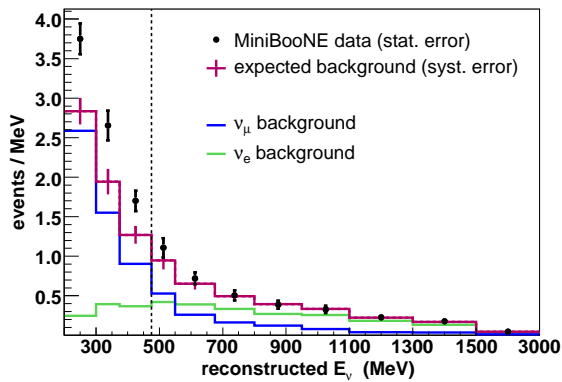}
\caption{ Reconstructed neutrino energy spectrum from MiniBooNE for
  data (black dots) and Monte Carlo (magenta line). The data is shown
  with statistical error bars and the Monte Carlo shows the expected
  non-oscillation background including systematic errors. A low energy
  threshold of 475\,MeV was applied to the analysis.}
  \label{MBenergy}
\end{figure}

The MiniBooNE collaboration decided to open their blinded data set in
stages, revealing more and more details about the data while still
blind to the oscillation result. During this unblinding procedure, it
was found that data and Monte Carlo expectation resulted in a bad
$\chi^2$ for both the oscillation and null hypothesis. The disagreement
was found to come from the lowest energy events. It was therefore
decided to move the energy threshold for the analysis to 475\,MeV,
while still showing the lower energy events to the public. The
resulting visible energy spectrum for data and Monte Carlo is shown in
Figure \ref{MBenergy}.

\begin{figure}[ht]
\centering
\includegraphics[width=80mm,clip]{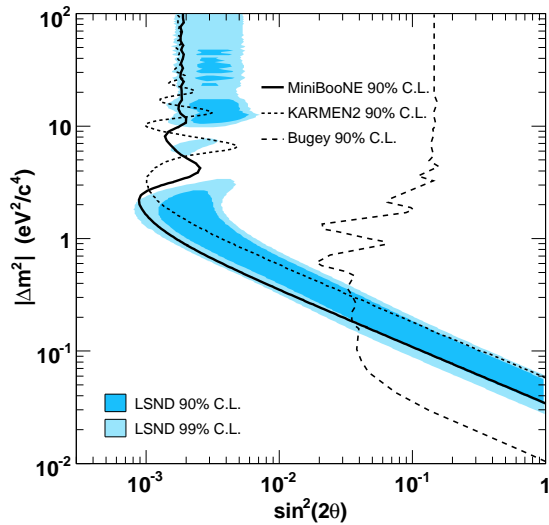}
\caption{The MiniBooNE 90\% exclusion curve is shown as a solid black
  line along with the curves from the KARMEN and Bugey experiments
  (dashed). The LSND allowed regions are shown in blue. MiniBooNE and
  LSND are incompatible at 98\% C.L. }
  \label{contour}
\end{figure}

Above the energy threshold, the data does not show a significant
excess and is consistent with no electron-neutrino appearance at the
1\,$\rm{eV}^2$ scale. MiniBooNE is therefore able to exclude a region
in the $\dm$--$\ssq$ parameter space. Figure \ref{contour} shows how
the MiniBooNE result compares with previous experiments Bugey, KARMEN
and LSND. It can be seen that MiniBooNE rules out large parts of the
LSND favoured region (shown in blue). Assuming that neutrinos and
antineutrinos oscillate in the same way and using a two-neutrino
approximation, LSND and MiniBooNE are incompatible at 98\,\% C.L.

Several potential sources for the reported excess of events below the
474\,MeV threshold have been discussed and are currently being
investigated by the collaboration. At the time of writing, no new
results on this issue were available.

\section{Summary}
Neutrino oscillations have been introduced and evidence and current
world knowledge on oscillation parameters from several experiments
have been reported. We reported the latest neutrino oscillation
results from the MINOS and MiniBooNE experiments, detailing aspects of
the MINOS \numu\ charged current, neutral current and \nue\ appearance
analysis as well as the MiniBooNE two-flavour oscillation analysis.

\begin{acknowledgments}
  The MINOS work reported here was supported by the US DOE; the UK
  STFC; the US NSF; the State and University of Minnesota; the
  University of Athens, Greece and Brazil's FAPESP and CNPq. MINOS
  would like to thank the Minnesota Department of Natural Resources,
  the crew of the Soudan Underground Laboratory, and the staff of
  Fermilab for their contributions to this effort.
  
  The author would like to thank the MiniBooNE collaboration and
  Rustem Ospanov for the material provided. Furthermore, we would like
  to thank the organisers of FPCP08 for the help provided.
\end{acknowledgments}


\end{document}